\documentstyle[emulateapj]{article}
\begin{document}
\submitted{Submitted February 27, 2001; published in ApJL}
\title{Abundances in the Uranium-Rich Star CS 31082-001}
\author{Y.-Z. Qian\altaffilmark{1} and G. J. Wasserburg\altaffilmark{2}}
\altaffiltext{1}{School of Physics and Astronomy, University of
Minnesota, Minneapolis, MN 55455; qian@physics.umn.edu.}
\altaffiltext{2}{The Lunatic Asylum,
Division of Geological and Planetary Sciences, California
Institute of Technology, Pasadena, CA 91125.}

\begin{abstract}
The recent discovery by Cayrel et al. of U in CS 31082-001 
along with Os and Ir at greatly enhanced abundances but with [Fe/H] 
$= - 2.9$ strongly reinforces the argument that there are at least 
two kinds of SNII sources for $r$-nuclei. 
One source is the high-frequency $H$ events responsible for 
heavy $r$-nuclei ($A>135$) but not Fe. The $H$-yields calculated from 
data on other ultra-metal-poor stars and the sun provide a template for 
quantitatively predicting the abundances of all other $r$-elements. In 
CS 31082-001 these should show a significant deficiency at $A < 135$ relative 
to the solar $r$-pattern. It is proposed that CS 31082-001 should have had
a companion that exploded as an SNII $H$ event.
If the binary survived the explosion,  this star should now have
a compact companion, most likely a stellar-mass black hole. Comparison of 
abundance data with predicted values and a search 
for a compact companion should provide a stringent test of the proposed
$r$-process model. The U-Th age determined by Cayrel et al. for
CS 31082-001 is, to within substantial uncertainties, in accord with the 
$r$-process age determined from solar system
data. The time gap between Big Bang and onset of normal star
formation only allows $r$-process chronometers to provide a 
lower limit on the age of the universe. 
\end{abstract}

\section{Introduction}
Recently Cayrel et al. (2001) reported discovery of U
along with a strong overabundance of the heavy $r$-process elements 
Os and Ir in the ultra-metal-poor (UMP, with [Fe/H]~$\approx -3$) 
halo star CS 31082-001. This star (hereafter the U-star) 
has [Fe/H]~$=-2.9$ and is similar in certain characteristics to another 
UMP halo star CS 22892-052 which has [Fe/H]~$=-3.1$ and also exhibits a 
large enhancement of heavy $r$-elements relative to Fe.
The possible use of $^{238}$U as a cosmochronometer was emphasized by 
Cayrel et al. (2001). Their data on the U-star also have important 
implications for stellar sites of the $r$-process and for  
nucleosynthetic yields of an $r$-process event.
In this Letter we argue that the observed abundances of Os, Ir, Th, and U
in the U-star are the result of the most frequent
Type II supernova (SNII) events and that the abundances of other 
$r$-elements can be predicted by a three-component model for 
abundances in metal-poor stars [Qian \& Wasserburg 2001 (QW)].
We also argue that the $r$-abundances in the U-star most likely reflect
the contamination of its surface by the ejecta from the SNII explosion of 
a previous massive binary companion. If the binary survived the
explosion, the U-star should now have a compact companion,
most likely a stellar-mass black hole. These inferences
could be directly tested by further observations of the U-star. We also
show that the U-Th age of halo stars provides 
an estimate for the ``local'' onset of $r$-process nucleosynthesis which  
only sets a lower bound on the age of the universe.  

The approach here follows that used by QW to
develop the three-component model for abundances in metal-poor stars 
based on the abundances of the radioactive nuclei $^{129}$I 
and $^{182}$Hf in the early solar system. These abundances require 
two distinct types of SNII if the $r$-process is associated with such 
events. The two types are the $H$ and $L$ events (Wasserburg, Busso, \& 
Gallino 1996; Qian, Vogel, \& Wasserburg 1998; Qian \& Wasserburg 2000). 
The $H$ events occur at a ``high'' frequency of 
$\approx (10^7\ {\rm yr})^{-1}$ in a standard reference mass of 
interstellar medium (ISM, mostly hydrogen) and are the major source of 
``heavy'' $r$-nuclei (with mass number $A>135$).
The $L$ events occur at a ``low'' frequency of 
$\approx (10^8\ {\rm yr})^{-1}$
and are the major source of ``light'' $r$-nuclei (with $A<135$). It was 
predicted that UMP stars would exhibit peculiar variations in abundances
relative to the solar $r$-pattern reflecting early contributions from 
the high-frequency $H$ events (Wasserburg et al. 1996). 
Evidence in support of this has been found
by Sneden et al. (2000) who showed that abundances of heavy 
$r$-elements from Ba ($A\sim 135$) and above in CS 22892-052 closely
follow the solar $r$-pattern, whereas the light $r$-elements around Ag
($A\sim 107$) are somewhat deficient. 

Data on Ba and Eu ($A\sim 151$)
for UMP stars (McWilliam et al. 1995; McWilliam 1998) show that there is 
a wide range in Ba/H and Eu/H for a fixed value of [Fe/H]~$\approx -3$.
This suggests that the $H$ events responsible for heavy $r$-nuclei
produce very little Fe [Wasserburg \& Qian 2000a (WQ)]. It was proposed that 
the (SNII) $H$ events are associated with black hole formation
(Qian et al. 1998). This is in accord with the scenario of Brown \& Bethe
(1994) and could explain the lack of Fe production in these events. The large
dispersion in Ba/H and Eu/H at [Fe/H]~$\approx -3$ was attributed by 
WQ to the rapid occurrence of high-frequency 
$H$ events. At [Fe/H]~$> -2.5$ an increasingly clear correlation 
between abundances of heavy $r$-elements and Fe is observed 
(e.g., Burris et al. 2000). 
This correlation would naturally be established as
products from both $H$ events and low-frequency Fe-producing $L$ 
events were mixed in the ISM. The Fe abundance 
resulting from an $L$ event is [Fe/H]$_L\approx -2.48$ (WQ). 

In addition to $H$ and $L$ events, an initial or prompt ($P$) inventory of 
elements is assumed to explain abundances 
of Fe and other associated elements in stars with [Fe/H]~$< -3$. The 
$P$-inventory was attributed to production 
by the first very massive ($\gtrsim 100\,M_\odot$) 
stars prior to formation of normal stars ($M\sim 1$--$60\,M_\odot$) and 
onset of regular SNII associated with the
$r$-process. It was argued that formation of normal stars could 
not occur until a ``metallicity'' corresponding to [Fe/H]~$\approx -3$ 
was achieved in the ISM (WQ). The $P$-inventory was
considered to represent the earliest stages of chemical evolution that 
need not be specifically associated with our Galaxy. Data on Ba at
$-4\lesssim {\rm [Fe/H]}<-3$ (McWilliam et al. 1995; McWilliam 1998;
Norris, Ryan, \& Beers 2001) 
indicate that the $P$-inventory of heavy $r$-elements is negligible.
As [Fe/H] for UMP stars reflect only the $P$-inventory of Fe
([Fe/H]$_P\approx -3$) and no $L$-contributions, we consider that 
abundances of heavy $r$-elements from Ba and
above in these stars only resulted from $H$ events. Observations by 
Sneden et al. (1996, 2000) and Westin et al. (2000) show that abundances 
of these elements in UMP stars very closely follow the solar $r$-pattern. 
Thus their $H$-yields can be directly calculated
from the solar $r$-abundances. Over the Galactic history of
$\approx 10$~Gyr before solar system formation 
$\approx 10^3$ $H$ events contributed to the solar abundances.
The $H$-yield of e.g., Os ($A\sim 190$) is then
$\log\epsilon_H({\rm Os})\approx\log\epsilon_{\odot,r}({\rm Os})-3=-1.66$
where $\log\epsilon({\rm E})\equiv\log({\rm E/H})+12$ for an element E and 
$\log\epsilon_{\odot,r}({\rm Os})=1.34$ (K\"appeler, Beer, \& Wisshak 1989; 
Arlandini et al. 1999) for the solar $r$-inventory of Os.
$H$-yields of heavy $r$-elements from Ba and above are given in Table 1. 

By comparing the observed abundances of Sr, Y, and Zr with those of the 
heavy $r$-elements in the UMP stars 
CS 22892-052 (Sneden et al. 2000) and HD 115444 (Westin et al. 2000), 
QW found that the Sr, Y, and Zr abundances in these
two stars received significant to dominant contributions from a common 
source, the $P$-inventory. 
The $\log\epsilon_P$ values for the  
$P$-inventory of Sr, Y, and Zr and the corresponding $H$-yields 
(represented by $\log\epsilon_H$) calculated by QW are given in Table 1. 
The observed Os abundance of $\log\epsilon({\rm Os})=-0.05$ in
CS 22892-052 (Sneden et al. 2000) indicates that this star received
contributions from $n_H\approx 41$ $H$ events. A slightly different value 
for $n_H$ is obtained by using the observed Eu 
abundance as in QW. The $H$-contributions to the
Sr, Y, and Zr abundances in CS 22892-052 are larger than the 
$P$-inventory. We assume that 
the observed abundances in CS 22892-052 for the light $r$-elements 
Nb, Ru, Rh, Pd, Ag, and Cd ($A=93$--116) are dominated by the 
$H$-contributions. Their $H$-yields are given in Table 1. 

$H$-yields of $r$-elements and the $P$-inventory of Sr, Y, and Zr are 
shown in Figure 1. As the U-star is a UMP
star with no $L$-contributions, $L$-yields calculated by QW are not given. 
In \S2 we discuss abundances of stable elements in the U-star based on 
the above three-component model. The abundances of 
$^{232}$Th and $^{238}$U in this star and implications for 
cosmochronology are discussed in \S3. 

\section{Abundances of Stable Elements in the U-star}
Data for the U-star clearly show an extremely large enhancement of heavy 
$r$-elements without any increase in [Fe/H] from the prompt value
[Fe/H]$_P\approx -3$. The observed Os abundance of 
$\log\epsilon({\rm Os})=0.49$ is higher than those in
CS 22892-052, HD 115444 ([Fe/H]~$=-2.99$), and HD 122563 ([Fe/H]~$=-2.74$;
Sneden et al. 1998) by 0.54, 1.04, and $>1.79$ dex, respectively.
Thus  observations of the U-star clearly confirm the existence of a type 
of $r$-process event that produces heavy $r$-nuclei but no Fe. This is 
in support of the hypothesis of the above three-component model that 
there must be two types of SNII sources for $r$-nuclei, one of which 
does not produce Fe.

Using $\log\epsilon_H({\rm Os})=-1.66$ we obtain from the observed 
Os abundance in the U-star that it had received
contributions from $n_H \approx 141$ $H$ events. The abundances 
of all other stable $r$-elements can be calculated  
from $n_H$ and the $H$-yields and the
$P$-inventory in Table 1. These calculated abundances are given in 
Table 1 and shown in Figure 1. While Sr, Y, and Zr have substantial 
contributions from the $P$-inventory, these are 
small compared with the contributions from $\approx 41$ $H$
events for CS 22892-052 or those from $\approx 141$ $H$ events for the U-star. 
Figure 1 also shows the solar $r$-pattern 
translated to fit the observed Os abundance in the U-star. 
The calculated abundances for the U-star show that it should be
deficient in the light $r$-elements from Sr to below Ba (particularly Y)
relative to the solar
$r$-pattern. Assuming the validity of the model, we consider 
the abundances predicted here for the U-star 
to be quantitative and reasonably reliable subject to uncertainties in 
the observational data from which the $H$-yields and
the $P$-inventory are derived. 
Absolute abundances relative to hydrogen in
this star will provide a direct and
rigorous test of the approach laid out here. We were informed by the
referee that these are under active study (Hill et al. 2001).
The essential conclusion from our results is that abundances in the 
U-star should reflect the relative yield pattern of an $H$ event in addition 
to simply exhibiting high $r$-process enrichments.

The high numbers of $H$ events for CS 22892-052 and the U-star
require discussion. These two UMP stars have
essentially the same [Fe/H] as the prompt value [Fe/H]$_P\approx -3$.
So they cannot have received contributions from any Fe-producing $L$
events. From the frequencies of $H$ and $L$ events in an average ISM,
the average fraction of $H$ events among all SNII is $q\approx 0.9$.
The probability for a standard reference mass of ISM to have a number
$n_H$ of $H$ events in a series is $q^{n_H}$. The number
$n_H\approx 41$ for CS 22892-052 corresponds to a small probability 
of $\approx 1.3\times 10^{-2}$ and the number $n_H\approx 141$
for the U-star to an extremely low probability of 
$\approx 3.5\times 10^{-7}$. It follows that the observed high enrichment
of $r$-elements in the U-star and possibly also CS 22892-052 
cannot be plausibly associated with many $H$ 
events randomly contaminating the ISM but requires a special source.

It takes $\approx 10^3$
$H$ events over $\approx 10$~Gyr to produce the solar $r$-process ratio 
(Os/H)$_{\odot,r}$ in a standard reference mass of ISM. 
For the present Galactic SNII rate of
$\approx (30\ {\rm yr})^{-1}$ corresponding to a total gas mass of
$\approx 10^{10}\,M_\odot$, the standard reference mass is
$\approx 3\times10^4\,M_\odot$. This is in accord with the typical mass
of ISM swept by an SNII remnant (e.g., Thornton et al. 1998). To explain
the observed Os abundance in the U-star by a special $H$ event 
contaminating the ISM would require either extremely large fluctuations in 
the $r$-process production by an $H$ event or wide variations in the amount 
of ISM that dilutes the $r$-process ejecta. It follows 
that the abundances in the U-star do not
represent a sample of the ISM with enormous enhancement in $r$-elements.
We propose that the U-star was once a binary companion to a massive
($\sim 10$--$60\,M_\odot$) star. The massive star exploded as an SNII
$H$ event and contaminated the U-star, providing a high enhancement in
its surface abundances of $r$-elements. If the $r$-process material
received from the SNII is mixed with $\sim 10^{-2}\,M_\odot$ of hydrogen
in the surface layer of the U-star, then only a fraction
$\sim 141(10^{-2}/3\times 10^4)\approx 5\times 10^{-5}$ of the 
$r$-process ejecta from an $H$ event is needed to give the observed 
enhancement of $r$-elements in this star. If the proposed binary survived
the SNII explosion, the U-star should now have a compact companion, most
likely a stellar-mass black hole. We were informed that the U-star has
a rather high proper motion for its distance. Long-term observations of 
its radial velocity should provide a definitive test of whether it is in a
binary. We note that a large 
overabundance of O, Mg, Si, and S associated with SNII explosions has
been observed in the binary companion to a possible black hole 
(Israelian et al. 1999). 

We have ignored the possibility that neutron
star mergers (NSMs) could be responsible for the heavy $r$-elements.
If NSMs were the source for such elements in the sun, the average enrichment
resulting from an NSM would be $\sim 3\times 10^3$ times that for
an SNII $H$ event (Qian 2000). To explain the observed Os abundance
in the U-star and the range of Os abundances in other UMP stars 
[a scatter in $\log\epsilon({\rm Os})$ of $>1.79$~dex] by NSMs would then 
require grossly variable amounts of ISM to mix with the ejecta. 
As NSMs play essentially no
role in Fe enrichment, the large scatter in abundances
of heavy $r$-elements should persist at [Fe/H]~$>-3$ if NSMs were the
sources for these elements (Qian 2000). However, the extensive data of 
Burris et al. (2000) on the heavy $r$-element Eu show that the scatter 
in $\log\epsilon({\rm Eu})$ is
$\approx 1$ dex at [Fe/H]~$\approx -2.5$ and decreases at higher [Fe/H].
Thus we do not consider that NSMs are the sources for the heavy 
$r$-elements, nor can they explain the observed $r$-abundances in the 
U-star. 

\section{$^{232}$Th and $^{238}$U Abundances in the U-star 
and Cosmochronology }
Abundances of $^{232}$Th and $^{238}$U in the U-star have been used
to determine the age of this star (Cayrel et al. 2001). If the $^{232}$Th 
and $^{238}$U observed in a star are the result of a single SNII $H$ 
event, the age equation for the star is
\begin{equation}
\label{age}
\left({^{238}{\rm U}\over{^{232}{\rm Th}}}\right)=
\left({Y_{238}\over Y_{232}}\right)
\exp\left[-\left({t_{\rm star}\over\bar\tau_{238}}- 
{t_{\rm star}\over\bar\tau_{232}}\right)\right],
\end{equation}
where $Y_{238}/Y_{232}$ is the relative yield of $^{238}$U to $^{232}$Th 
for the $H$ event, $\bar\tau_{238}=6.45$~Gyr and 
$\bar\tau_{232}=20.3$~Gyr are the lifetimes of the two
nuclei, and $t_{\rm star}$ is the time interval between the event and
observation. As emphasized by Cowan et al. (1999) and Goriely \& Clerbaux
(1999), the calculated age depends critically on the relative yields that
are very difficult to determine from {\it ab initio} $r$-process models.
The solar inventory of $r$-nuclei at the time of solar system formation
(SSF; 4.55 Gyr before the
present time) is the result of previous Galactic nucleosynthesis. Assuming 
that the rate of SNII production per hydrogen atom in the ISM is constant, 
we have
\begin{equation}
\label{sun}
\left({^{238}{\rm U}\over{^{232}{\rm Th}}}\right)_{\odot}^{\rm SSF}= 
\left({Y_{238}\over Y_{232}}\right)\left[\left({\bar\tau_{238}\over
\bar\tau_{232}}\right){1-\exp(-T_{\rm UP}/\bar\tau_{238})\over
1-\exp(-T_{\rm UP}/\bar\tau_{232})}\right],
\end{equation}
where $T_{\rm UP}$ is the total time of uniform $r$-process production
prior to SSF. The ratio $(^{238}{\rm U}/^{232}{\rm Th})_{\odot}^{\rm SSF}
=0.431$ (Anders \& Grevesse 1989) has an uncertainty of $<10\%$.
The term in square brackets in equation (\ref{sun}) ranges from 1 
(for $T_{\rm UP}=0$) to 0.318 ($T_{\rm UP}=\infty$). Thus there is
only an extremely restricted range in $Y_{238}/Y_{232}$
for a wide range in $T_{\rm UP}$. The relationship between 
timescales and long-lived and short-lived $r$-nuclei has been discussed 
extensively by Schramm \& 
Wasserburg (1970). An estimate of $Y_{238}/Y_{232}$ 
was first made by Burbidge et al. (1957) and later by 
Fowler \& Hoyle (1960) based on the number of progenitors
for $^{232}$Th and $^{238}$U. The estimated value of 
$Y_{238}/Y_{232}\approx 0.61\pm 0.06$ with guesstimated errors
has not been substantially improved upon over the past 
40 years (cf. Cowan et al. 1999; Goriely \& Clerbaux 1999). 
For these values of $Y_{238}/Y_{232}$ we obtain 
$T_{\rm UP}\approx 7.5\pm 2.5$~Gyr which
corresponds to a time of $\approx 12\pm 2.5$~Gyr since the onset of
Galactic $r$-process nucleosynthesis. 
The age of $12.5\pm 3$~Gyr for the U-star obtained by Cayrel et al. (2001) 
is for a single event. The value $Y_{238}/Y_{232}\approx 0.61$ corresponds
to an age of $\approx 11.4$~Gyr for this star. The estimated ages of the
U-star, the values of $T_{\rm UP}$
calculated from the solar system data for uniform $r$-process production
rates, and the timescale calculated by Fowler \& Hoyle 
(1960) for an exponential model are in general accord.

The values of $T_{\rm UP}$ are also consistent with the uniform
production period assumed above to calculate the number of SNII $H$
events contributing to the solar inventory. The $H$-yield of
$^{232}$Th can be calculated from
\begin{equation}
\label{th}
\left({^{232}{\rm Th}\over{\rm H}}\right)_{\odot}^{\rm SSF}= 
\left({^{232}{\rm Th}\over{\rm H}}\right)_Hf_H\bar\tau_{232}\left[
1-\exp\left(-{T_{\rm UP}\over\bar\tau_{232}}\right)\right],
\end{equation}
where $f_H\approx(10^7\ {\rm yr})^{-1}$ is the frequency of $H$ events.
For $T_{\rm UP}\approx 10$~Gyr we obtain $\log\epsilon_H(^{232}{\rm Th})
\approx -2.72$. A value of $\log\epsilon_H(^{238}{\rm U})\approx -2.89$ is 
similarly obtained. These yields are consistent with the observed Th and U
abundances in the U-star for an age of 14.5~Gyr (see Figure 1).

A deeper question is the relationship of ages obtained from 
$r$-process chronometers to the age of the universe. It has been argued 
above that onset of $r$-process nucleosynthesis was 
represented by rapid occurrence of $H$ events at [Fe/H]~$\approx -3$.
These events were preceded by nucleosynthesis in the first very massive 
stars that significantly produced only elements up to Zr. 
The timescale over which aggregation of matter proceeded until a 
metallicity corresponding to [Fe/H]~$\approx -3$ was achieved to permit
formation of normal stars is not established. In addition, 
assembly, disassembly, and reassembly of baryonic matter at various 
stages in earlier epochs of the universe possibly leave a large time 
interval prior to onset of SNII. From consideration of [Fe/H] 
in damped Ly$\alpha$ systems, it has been argued that there is a long
interval ($\sim 1$--5~Gyr) between Big Bang and  
onset of normal star formation in protogalaxies (Wasserburg \& 
Qian 2000b). Thus the ages obtained from $r$-process chronometers are 
substantially less than the age of the universe.

\acknowledgments
We would like to dedicate this paper to 
Willy Fowler and Fred Hoyle who might at sometime have 
enjoyed and engaged in these efforts. Support, interest, and provocation 
by Roger Blandford are greatly appreciated. We thank Tim Beers for a
thorough, insightful, and prompt review. We acknowledge Kris Davidson and
Roberta Humphreys for information on the proper motion of the U-star.
This work was supported in part by DOE grants DE-FG02-87ER40328 and 
DE-FG02-00ER41149 (Y.Z.Q.) and by NASA grant NAG5-4083 (G.J.W.), 
Caltech Division Contribution 8764(1075).

\figcaption{The $r$-process yields of an SNII $H$ event (dashed curve)
and the prompt inventory (dot-dashed curve) for the three-component 
model. The solid curve represents the abundances calculated 
(see Table 1) for the U-star by using the observed Os abundance to 
determine $n_H$. The Th and U abundances are 
calculated for an age of 14.5 Gyr. Filled circles with error bars indicate
data from Cayrel et al. (2001). The dotted curve marked ``Translated Solar''
represents the standard solar $r$-abundances (K\"appeler et al. 1989; 
Arlandini et al. 1999) shifted down to fit the Os data. 
We note that based on the data of Burris et al. (2001), QW have given 
revised solar $r$-abundances for Sr, Zr, and Ba.}

\begin{deluxetable}{ccrrr}
\footnotesize
\tablecaption{Abundances in the U-star, $H$-yields, and $P$-inventory}
\tablewidth{0pt}
\tablehead{
\colhead{Atomic Number ($Z$)}&\colhead{Element}&
\colhead{$\log\epsilon(Z)$}&
\colhead{$\log\epsilon_H(Z)$}&
\colhead{$\log\epsilon_P(Z)$}
}
\startdata
38&Sr&0.93&$-1.30$&0.13\\
39&Y&0.13&$-2.05$&$-1.05$\\
40&Zr&0.69&$-1.53$&$-0.13$\\
41&Nb&$-0.46$&$-2.61$&\nodata\\
44&Ru&0.54&$-1.61$&\nodata\\
45&Rh&$-0.26$&$-2.41$&\nodata\\
46&Pd&0.28&$-1.87$&\nodata\\
47&Ag&$-0.26$&$-2.41$&\nodata\\
48&Cd&0.19&$-1.96$&\nodata\\
56&Ba&0.63&$-1.52$&\nodata\\
57&La&$-0.07$&$-2.22$&\nodata\\
58&Ce&0.13&$-2.02$&\nodata\\
59&Pr&$-0.36$&$-2.51$&\nodata\\
60&Nd&0.25&$-1.90$&\nodata\\
62&Sm&$-0.05$&$-2.20$&\nodata\\
63&Eu&$-0.33$&$-2.48$&\nodata\\
64&Gd&0.15&$-2.00$&\nodata\\
65&Tb&$-0.55$&$-2.70$&\nodata\\
66&Dy&0.23&$-1.92$&\nodata\\
67&Ho&$-0.38$&$-2.53$&\nodata\\
68&Er&0.02&$-2.13$&\nodata\\
69&Tm&$-0.78$&$-2.93$&\nodata\\
70&Yb&$-0.07$&$-2.22$&\nodata\\
71&Lu&$-0.83$&$-2.98$&\nodata\\
72&Hf&$-0.46$&$-2.61$&\nodata\\
73&Ta&$ -1.21$&$-3.36$&\nodata\\
74&W&$-0.53$&$-2.68$&\nodata\\
75&Re&$-0.60$&$-2.75$&\nodata\\
76&Os&0.49&$-1.66$&\nodata\\
77&Ir&0.52&$-1.63$&\nodata\\
78&Pt&0.81&$-1.34$&\nodata\\
79&Au&$-0.05$&$-2.20$&\nodata\\
80&Hg&$-0.17$&$-2.32$&\nodata\\
81&Tl&$-0.65$&$-2.80$&\nodata\\
82&Pb&0.50&$-1.65$&\nodata\\
83&Bi&$-0.16$&$-2.31$&\nodata\\
90&$^{232}$Th&$-0.88$&$-2.72$&\nodata\\
92&$^{238}$U&$-1.72$&$-2.89$&\nodata\\
\enddata
\tablecomments{Column three give the calculated abundances in the U-star,
column four the $H$-yields, and column five the prompt inventory (values
for $Z>40$ are negligible). The $^{232}$Th and $^{238}$U abundances 
are calculated for an age of 14.5 Gyr.}
\end{deluxetable}


\begin{references}
\reference{}
Anders, E., \& Grevesse, N. 1989, \gca, 53, 197
\reference{}
Arlandini, C., K\"appeler, F., Wisshak, K., Gallino, R., Lugaro, M., 
Busso, M., \& Straniero, O. 1999, \apj, 525, 886
\reference{}
Brown, G. E., \& Bethe, H. A. 1994, \apj, 423, 659
\reference{}
Burbidge, E. M., Burbidge, G. R., Fowler, W. A., \& Hoyle, F. 1957,
Rev. Mod. Phys., 29, 547
\reference{}
Burris, D. L., Pilachowski, C. A., Armandroff, T. E., Sneden, C., 
Cowan, J. J., \& Roe, H. 2000, \apj, 544, 302
\reference{}
Cayrel, R., et al. 2001, \nat, 409, 691
\reference{}
Cowan, J. J., Pfeiffer, B., Kratz, K.-L., Thielemann, F.-K., Sneden, C.,
Burles, S., Tytler, D., \& Beers, T. C. 1999, \apj, 521, 194
\reference{}
Fowler, W. A., \& Hoyle, F. 1960, Ann. Phys., 10, 280
\reference{}
Goriely, S., \& Clerbaux, B. 1999, \aap, 346, 798
\reference{}
Hill, V., Plez, B., Cayrel, R., \& Beers, T. C. 2001, poster paper presented
at the Conference on Astrophysical Ages and Time Scales (Hilo, Hawaii)
\reference{}
Israelian, G., Rebolo, R., Basri, G., Casares, J., \& Martin, E. L. 1999,
\nat, 401, 142
\reference{}
K\"appeler, F., Beer, H., \& Wisshak, K. 1989, Rep. Prog. Phys., 52, 945
\reference{}
McWilliam, A., Preston, G. W., Sneden, C., \& Searle, L. 1995, 
\aj, 109, 2757
\reference{}
McWilliam, A. 1998, \aj, 115, 1640
\reference{}
Norris, J. E., Ryan, S. G., \& Beers, T. C. 2001, \apj, submitted
\reference{}
Qian, Y.-Z. 2000, \apj, 534, L67
\reference{}
Qian, Y.-Z., Vogel, P., \& Wasserburg, G. J. 1998, \apj, 494, 285
\reference{}
Qian, Y.-Z., \& Wasserburg, G. J. 2000, Phys. Rep., 333--334, 77
\reference{}
Qian, Y.-Z., \& Wasserburg, G. J. 2001, \apj, submitted (QW)
\reference{}
Schramm, D. N., \& Wasserburg, G. J. 1970, \apj, 162, 57
\reference{}
Sneden, C., Cowan, J. J., Burris, D. L., \& Truran, J. W. 1998, \apj, 496, 235
\reference{}
Sneden, C., Cowan, J. J., Ivans, I. I., Fuller, G. M., Burles, S., 
Beers, T. C., \& Lawler, J. E. 2000, \apj, 533, L139
\reference{}
Sneden, C., McWilliam, A., Preston, G. W., Cowan, J. J., 
Burris, D. L., \& Armosky, B. J. 1996, \apj, 467, 819
\reference{}
Thornton, K., Gaudlitz, M., Janka, H.-Th., \& Steinmetz, M. 1998,
\apj, 500, 95
\reference{}
Wasserburg, G. J., Busso, M., \& Gallino, R. 1996, \apj, 466, L109
\reference{}
Wasserburg, G. J., \& Qian, Y.-Z. 2000a, \apjl, 529, L21 (WQ)
\reference{}
Wasserburg, G. J., \& Qian, Y.-Z. 2000b, \apjl, 538, L99
\reference{}
Westin, J., Sneden, C., Gustafsson, B., \& Cowan, J. J. 2000, 
\apj, 530, 783
\end{references}
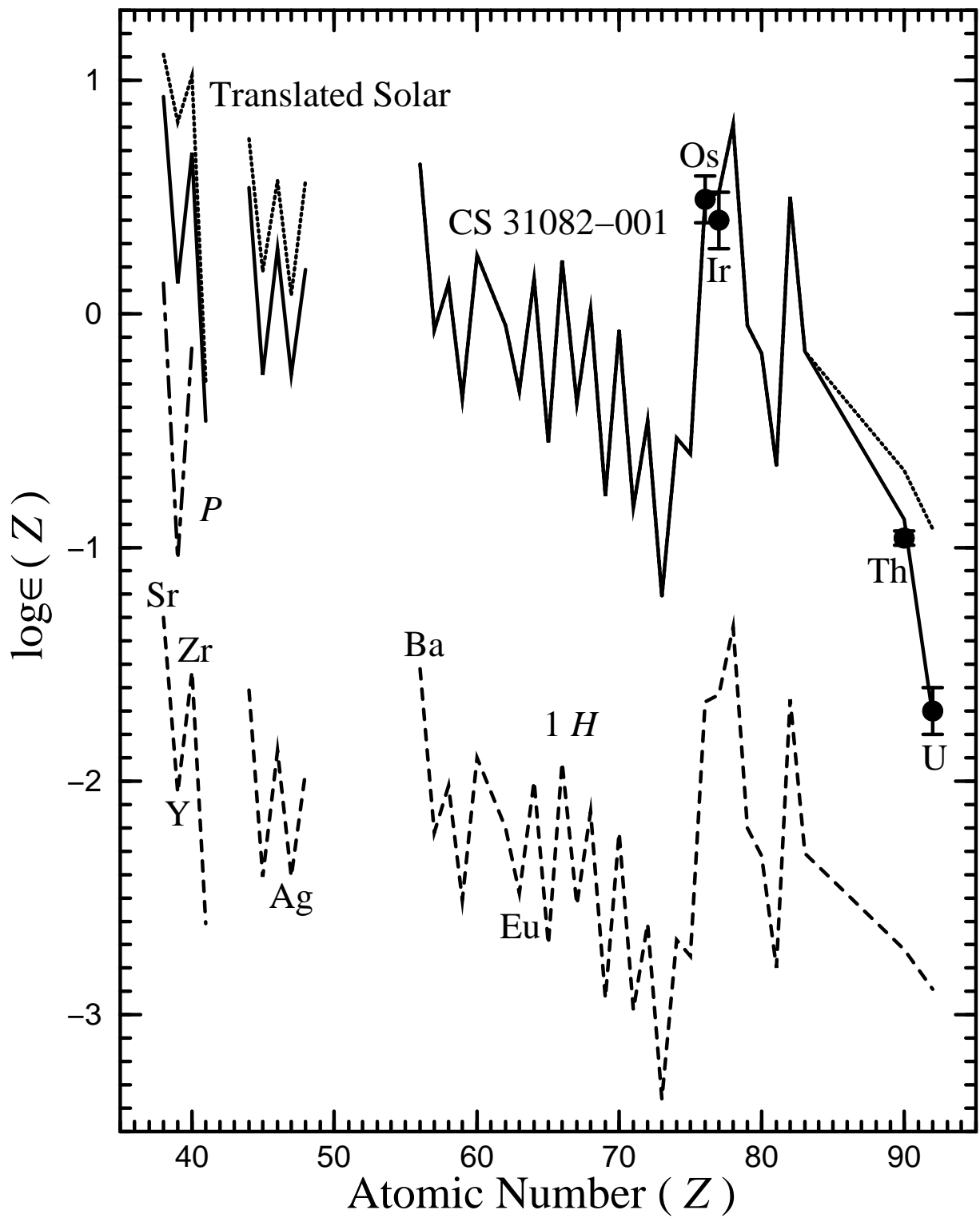
\end{document}